\documentstyle[12pt,titlepage]{article}

\advance\textheight by \topskip
\newcommand{\beq}{\begin{equation}}
\newcommand{\eeq}{\end{equation}}
\newcommand{\bea}{\begin{eqnarray}}
\newcommand{\eea}{\end{eqnarray}}
\newcommand{\beas}{\begin{eqnarray*}}
\newcommand{\eeas}{\end{eqnarray*}}
\newcommand{\beqs}{\begin{displaymath}}  
\newcommand{\eeqs}{\end{displaymath}}    

\newcommand{\me}{\mbox{}}

\newcommand{\G}{{\rm G}}

\begin{document}

\title{\bf
  The cosmological background in the Higgs scalar-tensor theory
  without Higgs particles
}
\author{
  H.~Frommert, H.~Schoor, and H.~Dehnen
  \\
  \rm E-Mail: \tt Hartmut.Frommert@uni-konstanz.de\\[1ex]
  \em Dept.\ of Physics, University of Constance\\
  \em P.O.Box 55 60 M 678, D-78434 Konstanz, Germany
}

\maketitle

\thispagestyle{empty}
\section*{Abstract}
The scalar background field and its consequences are discussed for the
Friedmann type cosmological solutions of the scalar-tensor theory of
gravity with the Higgs field of the Standard Model as the scalar
gravitational field.

\vfill

\noindent
\begin{description}
\item[PACs]: 4.20, 4.55, 12.10
\item[Keywords]:
  Higgs scalar-tensor theory;
  Higgs mechanism without Higgs particles;
  particle theoretical implications to gravitation theory;
  induced gravity;
  cosmological solutions
\end{description}

\clearpage
\pagenumbering{arabic}

\section{Introduction}

A scalar-tensor theory of gravity was developed by Brans and Dicke
\cite{bd61}
in order to introduce some foundation for the inertial mass as well as
the active and passive gravitational mass (i.e., the gravitational `constant')
of matter, by a scalar function determined by the distribution of all other
matter (i.e., particles) in the universe; the background of this is Mach's
principle and the principle of equivalence.

This introduction of mass by a scalar field can now be regarded as a somehow
prophetic approach, because in today's Standard Model of particle physics
the masses of the elementary particles are generated via the Higgs mechanism,
thus using also a scalar field, the Higgs field.
The scalar interaction mediated by the Higgs field was investigated by
Dehnen, Frommert, and Ghaboussi \cite{dfg90}.
They showed that any excited Higgs field%
  \footnote{
    The quanta of this excited Higgs field are the hypothetical Higgs
    particles.
  }
mediates an attractive scalar interaction%
  \footnote{
    This interaction is similar to gravity because it couples to the masses
    of the particles.
  }
of Yukawa type (i.e.\ short range)
between those particles which acquire mass by the corresponding symmetry
breaking (i.e.\ the fermions and the massive $W$ and $Z$ gauge bosons).
The Higgs field of particle physics can also serve as the scalar field in
a scalar-tensor theory of gravity, as was first proposed by Zee \cite{zee79}
and deeper investigated by Dehnen, Frommert, and Ghaboussi \cite{dfg92}.
In this theory, in addition to its role in the Standard Model to make the
particles massive, the scalar Higgs field also generates the gravitational
constant G, in the sense discussed by Adler \cite{adler82}
of generating an `induced' G from symmetry breaking.
Surprisingly however, if the Higgs field of the
$SU(3)\times SU(2)\times U(1)$ Standard Model of the elementary particles
is employed to generate G, the Higgs field looses its source,
i.e.\ can no longer be generated by fermions and gauge bosons
unless in the very weak gravitational channel.
Similar results were obtained independently by van der Bij \cite{vdb93}.
As Styp-Rekowski and Frommert \cite{sf96} have shown, the only physically
meaningful {\em static\/} solution of this theory is the trivial one
without any excited Higgs field present.
As the physical world is not static, the cosmological background is most
interesting, both on its own (e.g., for inflational scenarios; see 
Cervantes-Cota and Dehnen \cite{cd95})
and in order to have a fit for a more realistic
physical (i.e., dynamical) model configuration (e.g., galaxy).

The reader can find the whole formalism of this theory in Dehnen and
Frommert \cite{df93}.

\section{Friedmann type cosmology in the Higgs sca\-lar-tensor theory}

For the excited Higgs field $\varphi$, one obtains
the following homogeneous, covariant Klein-Gordon equation%
  \footnote{
    Throughout this paper we use
    $\hbar=c=1$ and the metric signature $(+---)$.
    The symbol $(\ldots)_{|\mu}$ denotes the partial,
    $(\ldots)_{\|\mu}$ the covariant derivative with respect to the
    coordinate $x^\mu$.
    For the cosmological discussion here, we also include the cosmological 
    constant $\Lambda$ which was omitted in the previous works.
  }
(see \cite{df93}):
\beq
\xi^{|\mu}_{\ \ \|\mu} + M^2 \xi
- {{4\over3} \Lambda \over1 + {4\pi\over3\alpha}}
= 0\ ,\ \ \
\xi = (1+\varphi)^2-1\ ,\ \ \
  \alpha\simeq 10^{33}
\label{eq:scf}
\eeq
where $M$ denotes the mass of the Higgs particles in this theory.  The field
equation for the metric as the tensorial gravitational field reads:
\bea
\lefteqn{
R_{\mu\nu} - {1 \over 2} R g_{\mu\nu} + {\Lambda\over 1+\xi} g_{\mu\nu}}
\nonumber \\*
&=& - {8\pi\G\over1+\xi} \Biggl[T_{\mu\nu}
+ {v^2\over4\left(1+\xi\right)} \left(\xi_{\vert\mu}\xi_{\vert\nu}
- {1\over2}\xi_{\vert\lambda}\xi^{\vert\lambda} g_{\mu\nu}\right)
+ V(\xi) g_{\mu\nu}\Biggr]
\nonumber \\*
&&{} - {1 \over 1+\xi} \left[\xi_{\vert\mu\Vert\nu}
- \xi^{\vert\lambda}_{\ \ \Vert\lambda} g_{\mu\nu} \right]\ .
\label{eq:gravf3chi}
\eea
with the Ricci tensor $R_{\mu\nu}$ belonging to the metric $g_{\mu\nu}$
and $T_{\mu\nu}$ as energy-momentum tensor of matter;
the Higgs potential is defined by:
\beq
V(\xi) = {3\over32\pi\G} M^2 \left(1+{4\pi\over3\alpha}\right) \xi^2
  \approx {3 M^2\over32\pi\G} \xi^2
\label{eq:pot}
\eeq
Because of the very large value of $\alpha$ (which is responsible for the
relative weakness of gravity as well as for the rather small Higgs mass in
this theory), we will neglect the $1/\alpha$ terms in equations
(\ref{eq:scf}) and (\ref{eq:pot}) in the following.

Here we look for solutions of the scalar field equation (\ref{eq:scf})
on the cosmological background of a Robertson-Walker metric,
defined by the line element (in isotropic spatial coordinates):
\beq
ds^2 = dt^2 - a(t)^2 {1\over 1 - (\varepsilon/4) r^2}
  \left(dr^2 + r^2 d\Omega^2\right)
\eeq
where, as usual, $\varepsilon=0,+1,-1$ corresponds to the spatially flat, the
closed, and the open model universe, respectively, and
$a(t)$ is the time dependent scale factor.
Moreover, we approximate the matter in cosmos by a perfect fluid,
characterized by its density $\varrho$ and pressure $p$ only, as usual,
and demand that $\varrho$ and $p$ as well as the scalar field $\xi$ are
functions of the time coordinate $t$ only (cosmological principle).
Then the scalar field equation (\ref{eq:scf}) becomes%
\footnote{
  $(\ldots)\dot{\me}:={\partial\over\partial t}(\ldots)$ denotes the time 
  derivative of $(\ldots)$.
}
\beq
\ddot\xi + 3{\dot a\over a}\dot\xi + M^2 \xi - 
{{4\over3}\Lambda 
} \equiv 0 \label{eq:scf0}
\eeq
while the nontrivial components of the Einstein equations are the
Friedmann equations:
\bea
{\dot a^2+\varepsilon\over a^2} - {\Lambda/3\over1+\xi}
&=& {8\pi\G\over3}{1\over1+\xi}
\left[\varrho + {v^2/8\over1+\xi}\dot\xi^2\right]
\nonumber
\\*
&&\me + {1\over1+\xi}\left[-{\dot a\over a}\dot\xi
  + {1\over4} M^2 \left(1+{4\pi\over3\alpha}\right) \xi^2\right]
\label{eq:fm1}
\\
2{\ddot a\over a} + {\dot a^2+\varepsilon\over a^2} - {\Lambda\over1+\xi}
&=& - {8\pi\G\over1+\xi} \left[p + {v^2/8\over1+\xi}\dot\xi^2\right]
\nonumber
\\*
&&\me - {1\over1+\xi} \left[\,\ddot\xi + 2 {\dot a\over a}\dot\xi -
{3\over4}
\left({M\over\hbar}\right)^2 \left(1+{4\pi\over3\alpha}\right) \xi^2
\right]\ .
\label{eq:fm2}
\eea

\noindent
These equations are augmented by the equation of continuity for the 
energy momentum tensor, which here reduces to
\beq                                                    
\dot\varrho + 3{\dot a\over a} (\varrho + p) =          
{\dot\xi\over 2\left(1+\xi\right)} (\varrho - 3p)\ .
\label{eq:cont}
\eeq                                                    

For the discussion of the cosmological background scalar field, one
notes first that, up to terms proportional to the Hubble constant
$H=\dot a/a$ and the cosmological constant $\Lambda$, the scalar field
equation (\ref{eq:scf0}) is solved by the periodic function
\beq
\xi = \xi_0 \cos\left[\omega\left(t-t_0\right)\right]\ ,\ \
\omega = \omega_M = M\ \ \left(= {M c^2\over\hbar} \right)
\label{eq:sol0}
\eeq
with a constant amplitude $\xi_0$ and period $\omega_M$, which is the 
Compton frequency belonging to the Higgs mass $M$, and thus {\em very}
large compared to the Hubble constant. One could expect modifications of
$\xi_0$ and $\omega_M$ for the exact solutions which are time dependent,
but change significantly only on the cosmological time scale, so that 
for all non-cosmological considerations equation (\ref{eq:sol0}) should 
be a good approximation.

For the following more detailed discussion, we note that the term with the
cosmological constant $\Lambda$ in the scalar field equation 
(\ref{eq:scf0}) is even small compared to the one containing
the Hubble constant, and can be absorbed in $\xi$ as a small
additive contribution. Therefore, we can restrict ourselves here
to discuss the equation for $\Lambda=0$ only, which reads:
\beq
\ddot\xi + 3{\dot a\over a}\dot\xi + M^2 \xi = 0
\label{eq:sfc1}
\eeq
This equation may be simplified with the ansatz:
\beq
\xi(t) =: a^{-3/2} u(t)
\eeq
Herewith the field equation (\ref{eq:sfc1}) reads:
\beq
\ddot u + 
  \left[ M^2 - {3\over 4} \left({\dot a\over a}\right)^2
    - {3\over 2} {\ddot a\over a} \right] u = 0\ .
\label{eq:sfc2}
\eeq
Up to terms of order $H^2$ and $\ddot a/a = - q H^2$
(with the cosmological acceleration parameter%
\footnote{
  $q$ is half the density parameter $\Omega$ in standard
  Friedmann cosmology without cosmological $\Lambda$.
}
$q$),
the solution of
this equation is identical to equation (\ref{eq:sol0}).
Using the cosmological parameters, the Hubble constant $H$ and the
acceleration parameter $q$, equation (\ref{eq:sfc2}) reads
\beq
\ddot u + \left[ M^2 + {3\over 2} \left( 2q-1 \right) H^2 \right] u = 0\ .
\label{eq:sfc3}
\eeq
For times small compared to the Hubble time $1/H$, this equation
should be approximately solved by the solutions (\ref{eq:sol0}) of the
approximated equation (\ref{eq:scf0}), so that:
\beq
u = u_0 \cos\left[M\left(t-t_0\right)\right]\ ,\ \
\xi = u_0 a^{-3/2} \cos\left[M\left(t-t_0\right)\right]
\label{eq:u-eq}
\eeq
This solution is accurate to one order more (i.e., to the second order)
in $H$, or considered time differences compared to the Hubble time, than
solution (\ref{eq:sol0}).
Some higher acuracy can be obtained by inserting the current values of the
function-valued parameters $H$ and $q$ into equation (\ref{eq:sfc3}):
\beq
\ddot u + \left[ M^2 + {3\over 2} \left( 2q_0-1 \right) {H_0}^2 \right] u = 0
\ ,\ \
q_0 = q(t_0) = const\ ,\ \ H_0 = H(t_0) = const\ ,
\label{eq:sfc4}
\eeq
which is solved by
\bea
u &=& u_0 \cos\left[
  \sqrt{1+{3\over2}(2q_0-1)\left({H_0\over M}\right)^2} M \left(t-t_0\right)
            \right]\ ,
\nonumber\\
\xi &=& u_0 a^{-3/2} \cos\left[
  \sqrt{1+{3\over2}(2q_0-1)\left({H_0\over M}\right)^2} M \left(t-t_0\right)
            \right]\ .
\label{eq:sol}
\eea
This solution differs from the above, equation (\ref{eq:u-eq}), by a slightly
different oscillation frequency, deviating from that in equation
(\ref{eq:u-eq}) (which is the Compton frequency corresponding to the
Higgs mass $M$) by a correction of the order given by the square of the
ratio%
\footnote{
  The value of this ratio can be estimated, taking into account that the
  Higgs mass in this theory is smaller than the usual one by a factor of
  about $2.5\cdot 10^{16}$, or at least about
  $2.5\cdot 10^{-6} eV/c^2$,
  corresponding to a Compton time of
  $\hbar/Mc^2\approx 2.6\cdot 10^{-10} s$.
  This must be compared to the Hubble time of 13 billion years (assuming
  $H_0=75 km/(s\>Mpc)$), so that
  $$
  ({H_0 \over M})^2 \approx 4 \cdot 10^{-55}
  $$
}
 of the two characteristic times relevant here, the Compton time $1/M$
corresponding to the Higgs mass $M$, and the Hubble time $1/H_0$: $(H_0/M)^2$.
The smallness of this value can already be seen by estimating the corrective
term in the field equation (\ref{eq:sfc4}), or the frequency in
(\ref{eq:sol}), using the second Friedmann equation (\ref{eq:fm2}), which
yields, neglecting the scalar field $\xi$:
$$
(2q-1) H^2 = {\varepsilon\over a^2} + 8\pi\G p + \Lambda;
$$
i.e., it is determined by the largest of its three terms:
If $\varepsilon=\pm1$ (non-flat case), baryonic
matter dominates ($p\ll \varrho$) and $\Lambda\ll a^{-2}$, it is essentially
given by $\pm 1/a^2$, while for the flat case, either the weak matter
pressure $p$ or the cosmological constant $\Lambda$ determines this
correction.
In view of this small value, one has to consider that for a self-consistent
solution, the contributions of the scalar field in the Friedmann equations
(\ref{eq:fm1}) and (\ref{eq:fm2}) must be taken into account, and it is not
guaranteed that these are smaller than the deviations discussed here.
This will be important for attempts to iterate cosmological solutions in
this theory.

One may also discuss the {\em exact\/} solutions. 
A self-consistent, exact and simultaneous solution of the relevant 
equations (\ref{eq:fm1},\ref{eq:fm2},\ref{eq:cont},\ref{eq:sfc1})
cannot be given analytically.
However, it can be obtained numerically to some approximation,
which was done by Cervantes-Cota and Dehnen \cite{cd95},
and is of particular
interest, especially in the context of possible inflation scenarios. 
Here we discuss analytic background solutions, which are obtained if 
a Friedmann solution is given as external field.  Then it is possible to
solve equation (\ref{eq:u-eq}) analytically for each given ansatz 
for $a(t)$. 
This is presented in the following for the two simplest cases, where
it is possible to give the exact solutions
(also mentioned in \cite{cd95}
as the limiting cases for the inflationary cosmology):
\begin{enumerate}
\item the spatially flat Friedmann universe with $\Lambda=0$,
\item the empty universe with cosmological constant $\Lambda$.
\end{enumerate}
One may hope that these background solutions are of interest in 
considerations where cosmology plays the role of a background, i.e.\
the dynamics of galaxies or clusters of galaxies.

\subsection{The spatially flat Friedmann universe with $\Lambda=0$}

In this case, we have the following time evolution of the scale 
parameter $a(t)$:
\beq
a(t) = A t^{2/3}
\eeq
and therefore
\bea
H(t) &=& {\dot a\over a}\ \ =\ \ {2\over 3 t}
\\
{\ddot a\over a} &=& - {2\over 9 t^2}\ \ \ \
\left(q\ \ =\ \ {1\over2}\right)
\eea
One may notice that this ansatz leads to the following equation for 
$\xi$:
\beqs
\xi = a^{{-3/2}} u = A^{{-3/2}} t^{{-1}} u
\eeqs
The differential equation (\ref{eq:sfc2}) for $u$
simplifies exactly to
\beq
\ddot u + M^2 u = 0\ ,
\eeq
which is identical to the equation without $H_0$, and has the 
exact solution
\beq
u(t) = u_0 \cos\left[M\left(t-t_0\right)\right]
\eeq
or
\beq
\xi = A^{{-3/2}} u_0 t^{-1} \cos\left[M\left(t-t_0\right)\right]
  = \xi_0 t_0 {\cos\left[M\left(t-t_0\right)\right]\over t}\ ,
\label{eq:sol1}
\eeq
where the relation for the amplitude at present time, 
$\xi_0 = A^{{-3/2}} u_0/t_0 = {a_0}^{{-3/2}} u_0$,
was used. To see the approximate constancy of the
amplitude over times small compared to the Hubble time (or world age),
one may expand the time $t$ as $t = t_0 + T$.
Then the solution (\ref{eq:sol1}) can be rewritten as
\beq
\xi = \xi_0(t) \cos\left[M\left(t-t_0\right)\right]\ ,\ \ 
\mbox{\rm where}\ \ 
\xi_0(t) = {\xi_0\over1+T/t_0} 
\approx \xi_0\left(1 - {3\over2} H_0 T\right)
\label{eq:xi0}
\eeq

\subsection{The empty universe with non-vanishing $\Lambda$
    (de Sitter universe)}

As a second ansatz, we investigate the case of the matter free universe 
with nonvanishing $\Lambda$ (matter influence neglegible compared with
that of the cosmological constant).
Then we have
\beq
H = {\dot a\over a} = \sqrt{\Lambda/3} = const\ ,\ \
a=a_0 e^{{H\left(t-t_0\right)}}\ ,\ \
q=-1\ .
\eeq
With this ansatz, the differential equation (\ref{eq:sfc2}) for $u$ 
takes the form
\beq
\ddot u + \left(M^2 - {9\over4} H^2\right) u = 0\ ,
\eeq
which has the solution
\beq
u = u_0 \cos\left[\sqrt{M^2-{9\over4}H^2}\left(t-t_0\right)\right]
\eeq
or
\beq
\xi = a^{{-3/2}} u = \xi_0 e^{{-{3\over 2} H\left(t-t_0\right)}}
  \cos\left[\sqrt{M^2-{9\over4}H^2}\left(t-t_0\right)\right]
\eeq
Again, with $t=t_0+T$, the time dependent amplitude $\xi_0(t)$ of this
solution (see (\ref{eq:xi0})) can be expanded in orders of $T/t_0$ or $Ht$,
and has the same form as above up to the first order:
$\xi_0(t)=\xi_0\left(1-(3/2)HT\right)$

\section{Remarks on the exact solutions}

The scalar field equation (\ref{eq:sfc2}) or (\ref{eq:sfc3}), in view of
the approximate solutions (\ref{eq:u-eq}) or (\ref{eq:sol}), may be rewritten
with the ansatz $u=u_0(t) \cos(\beta(t))$, where
$\beta(t) = M \int^t dt' \sqrt{1+\kappa(t')}$, so that
$\dot\beta = M \sqrt{1+\kappa(t)}$
($\kappa \ll 1$ if $H/M\ll 1$, $t_H\gg\hbar/(Mc^2)$).
Then one obtains:
\bea
\ddot u_0 + \left[- M^2\kappa + {3\over4} (2q-1) H^2 \right] u_0 &=& 0
\label{eq:xsol1}
\\
\dot\beta {u_0}^2\ \ =\ \ M \sqrt{1+\kappa} {u_0}^2\ \ =\ \ const &=:& C
\label{eq:xsol2}
\eea
The second of these equations, (\ref{eq:xsol2}), is e.g. solved by
$u_0=const$ and $\dot\beta=const$ or $\kappa(t)=const$; this is possible
if $q$ and $H$ in (\ref{eq:xsol1}) are taken as constants.
According to the remarks above, the deviation of $u_0$ and $\dot\beta$
or $\kappa(t)$ from constancy can be expected to be small, at least for
the large Hubble times considered here.
One may rewrite (\ref{eq:xsol1}) by evaluating (\ref{eq:xsol2}) in a
single nonlinear differential equation for $u_0(t)$:
\beq
\ddot u_0 + \left[M^2 + {3\over4} (2q-1) H^2\right] u_0
- {C^2\over {u_0}^3} = 0
\label{eq:xsol3}
\eeq
In view of the very small values of the corrections to the approximate
solution (\ref{eq:u-eq}) or (\ref{eq:sol}) we will not further discuss this
equation here.

\section{Effective Einstein equations with an averaged cosmological
         scalar field}

Having on hand that the scalar field solution for the cosmological 
background is given by a rapidly oscillating function, which is overlayed
over an amplitude which changes at cosmological time scales only, it is
of interest to take a new look at the basic field equations of the
theory.
With the scalar field amplitude given by
\beq
\xi_0 = a(t)^{-3/2} u_0 = \xi_0(t)
\eeq
where $u_0=u_0(t)$ must fulfill the differential equation
(\ref{eq:xsol1}) or (\ref{eq:xsol3}),
the time averaged%
\footnote{
  The averaged terms are obtained by Taylor-expanding the factors
  corresponding to the $1+\xi$ denominators, so that power series of
  $\xi$ are obtained, and then taking the time average given by the
  formula
  $$
  \langle F(t)\rangle = \langle F \rangle (t_0)
  = {1\over 2\pi} \int_{t_0}^{t_0+T} F(t') dt'
  $$
  where $T=1/\omega$ is the oscillation period, which must be small
  compared to the age of the universe, $t_0$.
  This average is a time function which varies slowly over cosmological
  time scales.
}
(over one period of the scalar oscillation)
Einstein equations take the form (``effective" Einstein equations):
\bea
\lefteqn{
  R_{\mu\nu} - {1\over2} R g_{\mu\nu} + 
  \left\langle{\Lambda\over1+\xi}\right\rangle g_{\mu\nu}\ \ =
}
\nonumber
\\
&=&
  R_{\mu\nu} - {1\over2} R g_{\mu\nu} + 
  {\Lambda\over\sqrt{1-{\xi_0(t)}^2}} g_{\mu\nu}
\nonumber
\\
&=& \left\langle
  - {8\pi\G\over1+\xi} \left[
       T_{\mu\nu} 
       + {v^2\over4\left(1+\xi\right)}\left(
           \xi_{,\mu}\xi_{,\nu}
          -{1\over2}\xi_{,\lambda}\xi^{,\lambda}g_{\mu\nu} 
         \right) 
       + V(\xi) g_{\mu\nu}
     \right]\right.
\nonumber\\*
&&
\left.\me
  - {1\over1+\xi} \left(
      \xi_{,\mu;\nu}-{1\over2} \xi_{,\lambda}^{\ ;\lambda} g_{\mu\nu}
    \right)
\right\rangle
\nonumber
\\
&\stackrel{!}{=}&
  - {8\pi\G\over\sqrt{1-{\xi_0(t)}^2}} T_{\mu\nu}
  - \delta_\mu^0 \delta_\nu^0 M^2 
     \left({1\over\sqrt{1-{\xi_0(t)}^2}}-1\right)
     \left(1-{2\pi\over\alpha}\right)
\nonumber
\\
&&\me + {M^2 \over 4} g_{\mu\nu} \left({1\over\sqrt{1-{\xi_0(t)}^2}}-1\right)
\eea

\noindent
This means that the scalar field leads to
\begin{itemize}
\item a (time dependent) effective gravitational ``constant"
  \beq
  \G_{\rm eff} = {\G\over\sqrt{1-{\xi_0(t)}^2}} > \G
  \eeq
\item a correction factor and a negative (attractive) contribution to the
  cosmological constant, or function, which becomes effectively:
  \beq
  \Lambda_{\rm eff} = {\Lambda\over\sqrt{1-{\xi_0(t)}^2}}
    - {M^2\over4} \left({1\over\sqrt{1-{\xi_0(t)}^2}}-1\right)
  \eeq
\item as well as a positive effective ``energy (or mass) density of the
  vacuum", given by
  \beq
  T^{{\rm vac}}_{\mu\nu}
  = \delta_\mu^0 \delta_\nu^0 \varrho_{\rm vac}
  = {\delta_\mu^0 \delta_\nu^0 M^2 \over 8\pi\G_{\rm eff}}
     \left({1\over\sqrt{1-{\xi_0(t)}^2}}-1\right)
     \left(1-{2\pi\over\alpha}\right)
  \eeq
\end{itemize}
As $\xi_0(t)$ changes over cosmological time scales only, it may be
regarded as constant in a first approximation unless cosmological 
aspects are discussed for themselves. This may be of particular interest
for the dynamics of galaxies (e.g., rotation curves) and galaxy 
clusters, i.e.\ the dark matter problem.
The calculation of the rotation curves for some disk galaxy models is
in preparation.

A possible limit for the time variations of $\xi_0(t)$ may be found from
geophysical or solar system results.


\end{document}